\documentclass[a4paper,11pt]{article}
\pdfoutput=1

\usepackage{jinstpub}
\usepackage{verbatim}
\usepackage{lineno}
\usepackage[T1]{fontenc}	
\usepackage[utf8]{inputenc}	
\usepackage[english]{babel}	
\usepackage{graphicx}
\usepackage{xcolor}

\usepackage[numbers,sort&compress]{natbib}

\title{Resolution enhancement with light/heat decorrelation in CUPID-0 bolometric detector}

\author[a,b,1]{M.~Beretta,\note{Corresponding author.}}      
\author[c]{L.~Cardani,}      
\author[c]{N.~Casali,}
\author[a,b]{L.~Gironi,}
\author[a,b]{L.~Pagnanini,}
\author[c,d]{F.~Bellini,}
\author[a,b]{C.~Brofferio,}
\author[a,b]{S.~Capelli,}
\author[a,b]{D.~Chiesa,}
\author[e,f]{S.~Di Domizio,}
\author[g,h]{L.~Pattavina,}
\author[a,b]{M.~Pavan,}
\author[g]{S.~Pirro,}
\author[a,b]{S.~Pozzi,}
\author[b]{E.~Previtali,}
\author[g,i]{C.~Rusconi,}      
\author[c]{C.~Tomei,}      
\author[c]{M.~Vignati}

\affiliation[a]{Dipartimento di Fisica - Universit\`{a} di Milano - Bicocca, 20126 Milano, Italy} \affiliation[b]{INFN - Sezione di Milano-Bicocca, 20126 Milano, Italy} 
\affiliation[c]{INFN - Sezione di Roma, 00185 Roma, Italy}
\affiliation[d]{Dipartimento di Fisica, Sapienza Universit\`{a} di Roma, 00185 Roma, Italy}
\affiliation[e]{Dipartimento di Fisica, Universit\`{a} di Genova, Genova I-16146 - Italy}
\affiliation[f]{INFN - Sezione di Genova, Genova I-16146 - Italy}
\affiliation[g]{INFN Laboratori Nazionali del Gran Sasso, 67100 Assergi (AQ), Italy}
\affiliation[h]{Gran Sasso Science Institute, 67100 L'Aquila, Italy}
\affiliation[i]{Department of Physics and Astronomy, University of South Carolina, Columbia, SC 29208, USA}

\emailAdd{mattia.beretta@mib.infn.it}

\abstract{The CUPID-0 experiment searches for neutrinoless double beta decay ($0\nu\beta\beta$) using the first array of enriched Zn$^{82}$Se scintillating bolometers with double (heat and light) read-out. To further enhance the CUPID-0 detector performance, the heat-light correlation has been exploited to improve the energy resolution. Different decorrelation algorithms have been studied and the best result is the average reduction of the full width at half maximum (FWHM) energy resolution to $(90.5\pm0.6)~\%$ of its original value , corresponding to a change from $\text{FWHM}=(20.7\pm0.5)~\text{keV}$ to  $\text{FWHM}=(18.7\pm0.5)~\text{keV}$ at the 2615~keV $\gamma$ line.}

\begin{document}
\maketitle
\flushbottom

\section{Introduction}
In the current framework of particle physics, research is directed towards the investigation of physical phenomena not predicted by the Standard Model. In this landscape, the search for neutrinoless double beta decay ($0\nu\beta\beta$) is of great importance, since its discovery would shed new light on neutrino phenomenology. The experimental signature for this process is a peak in the summed energy spectrum of the electrons at the transition energy of the decay ($Q_{\beta\beta}$), that must be identified above the background \cite{CHALLENGE_DBD}. Presently, only limits on the $0\nu\beta\beta$ half life ($T_{1/2}^{0\nu}$) have been measured for different isotopes, with values in the range $10^{24} - 10^{26}$~y \cite{0NUBB_REV_2016,BetaBetaVal}. The extremely low expected rates force the design of detectors capable of reaching low levels of background and good full width at half maximum (FWHM) energy resolution  at $Q_{\beta\beta}$.

To accomplish this goal, the CUPID-0 detector exploits the technique of scintillating bolometers, combining the high efficiency and energy resolution of bolometric detectors to the background rejection capabilities of scintillators \cite{Artusa2016}. The detector is made of 26 cylindric ZnSe crystals of different dimensions, each of which coupled to two germanium bolometers operated as light detectors (LDs) \cite{CUPIDDETECTOR}. A particle interaction inside the crystal causes both a temperature increase (heat signal) and a light emission due to the scintillation of ZnSe, detected by the light detectors (light signal). CUPID-0 allowed to set the most stringent lower limit on the half life of $^{82}Se$ $0\nu\beta\beta$ $T_{1/2}^{0\nu} > 2.4\cdot10^{24}$~yr (90~\% credible interval) \cite{CUPIDPRL}, proving the goodness of this technique. 
\\
To further increase the capability of this detector, the correlation between heat and light amplitude has been investigated. As reported in \cite{CdWO4CORR} and \cite{ZnSEDecorr}, in bolometers with dual readout the light/heat correlation can be used to correct the energy resolution over all the energy spectrum, improving the performances of the detector. This is a crucial issue in calorimetric experiment searching for $0\nu\beta\beta$, since better resolving detectors have higher sensitivity \cite{0NUBB_REV_2016}.
\\
In the next sections the method used to exploit the light/heat correlation will be discussed and the obtained results will be presented, in terms of FWHM reduction.
\section{Choice of the decorrelation method}

\begin{figure}[htbp]
\centering
\includegraphics[width=0.6\textwidth]{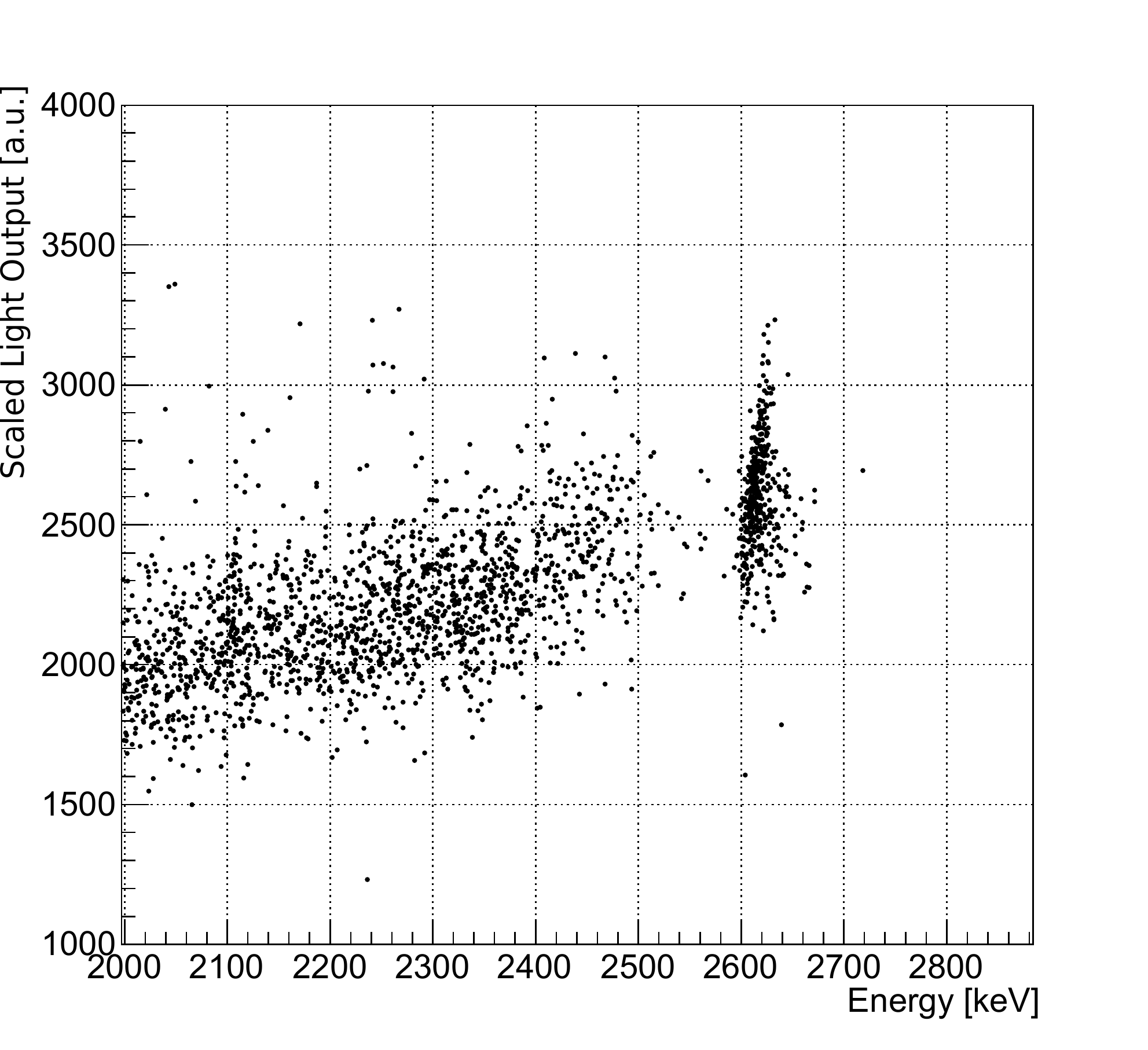}
\caption{Light/Heat scatter plot. The 2615~keV line clearly shows the correlation between the two variables. The resolution difference between heat and light channel can be appreciated. The heat axis was calibrated using the most prominent gamma peaks of a $^{232}$Th source (including the 2615~keV line shown in the plot). The light axis was re-scaled by assigning the nominal energy of the 2615~keV line to the corresponding scintillation peak.}
\label{fig:ScatLH}       
\end{figure}

In figure \ref{fig:ScatLH}, the light/heat scatter plot for one of the light detectors and its corresponding ZnSe calorimeter is reported. 
The events belonging to the 2615~keV $\gamma$ line clearly show the correlation between the two variables. ZnSe Crystals are characterized by a positive correlation between light and heat, while a negative correlation is usually present in other scintillating crystals \citep{CdWO4CORR}. No explanations exist for this effect, and such phenomenon contributes in worsening the energy resolution of the heat channel. A decorrelation method would enhance this parameter, already characterized by better signal to noise ratio with respect to the light signal.

The correlation between two variables can be corrected both by calculating a new variable combining light and heat information \cite{COMBBALZ} or by finding a geometrical transformation minimizing the resolution. Given the geometrical aspect of the correlation (see figure \ref{fig:ScatLH}), the optimal transformation to be applied is a counterclockwise rotation of the light/heat scatter plot. In the former method the corrected variable is calculated using standard deviation and correlation of the light/heat distribution, evaluated analytically \cite{ZnSEDecorr}. In the latter method the optimal rotation angle is chosen trying different possible values with an iterative algorithm.
\\
The two methods are mathematically equivalent. With the proper transformations, in fact, a rotation in the light/heat plane can be expressed as a weighted sum of heat and light variables. As a consequence, the minimization can be performed choosing the optimal weight analytically or the optimal angle numerically. Nevertheless, the practical application of these two strategies has different constraints, due to their different hypothesis. The analytical method is less computationally expensive than the second one, since it finds a complete expression for the weighting factor minimizing the resolution. This expression, reported in \cite{ZnSEDecorr}, depends on the standard deviations of heat and light distribution, as well as on the correlation factor between these two variables. The first aspect to be considered is that this derivation lies on the hypothesis of gaussian distributions both for heat and light. When these assumption is violated, the derivation of this optimal value can fail. Moreover, all the parameters of the optimal weight have to be calculated numerically from the heat/light distributions, like the one in figure \ref{fig:ScatLH}. Such evaluations are affected by uncertainties related to the finite resolution of the detector and to the non-flat background. Moreover,the poor signal to noise ratio of the light channel makes the evaluation of the standard deviation for this channel troublesome. As it is shown in figure \ref{fig:ScatLH}, in fact, the distribution along the light axis is more scattered than the distribution along the energy axis. Moreover the optimal weight is by construction dependent on energy. As stated in \cite{ZnSEDecorr}, it must show, at least, a little dependence on energy. As a consequence, the analytical evaluation must be performed for different energies and a fit of the obtained parameters with a polynomial is needed to extrapolate the correction at every energy.
\\
On the other hand, finding the optimal rotation which minimizes the heat resolution is based on the observation of changes in the heat channel resolution only. As a consequence, the poor signal to noise ratio of the light channel does not affect its performance. No assumption is made on the shape of the distributions used in this transformation, except for the existence of a correlation. In section \ref{Sec:StudiedAlg} a comparison between the results of analytic and numeric method will be presented, fully motivating the choice of the numeric approach.
\\
In CUPID-0 the ZnSe crystals are interleaved with germanium light detectors, in such a way that each ZnSe crystal faces two LDs: one above (Top) and one below (Bottom) \cite{CUPIDDETECTOR}.
Each LD has SiO anti-reflective coating (ARC) one the lower face, while the other one is bare.
Given the structure of the CUPID-0 tower, for each crystal the Top light detector has the ARC, while the Bottom detector has not. As a consequence, each heat signal generates two light signals, one for each LD. The presence of two different light signals results in two possible correlated quantities to be exploited for the resolution enhancement. In particular, the optimization can be performed both independently for the two light amplitudes or by combining these two informations. Both these investigations have been performed in the presented work, covering the possible combinations.
\\
The data used in this study belong to different DataSets (DSs), defined as periods of data-taking  approximately one month long started and finished with a $^{232}$Th calibration \cite{ANALISI_CUPID}. For each DS, the amplitude correction is performed on all the 26 heat channels of the detector, obtaining the resolution enhancement for each ZnSe crystal. Once the rotation angles for each channel are defined, all the informations are combined to evaluate the performances achieved in each DS.

\section{Studied algorithms for decorrelation} \label{Sec:StudiedAlg}

\begin{figure}[htbp]
\centering
\includegraphics[width=0.6\textwidth]{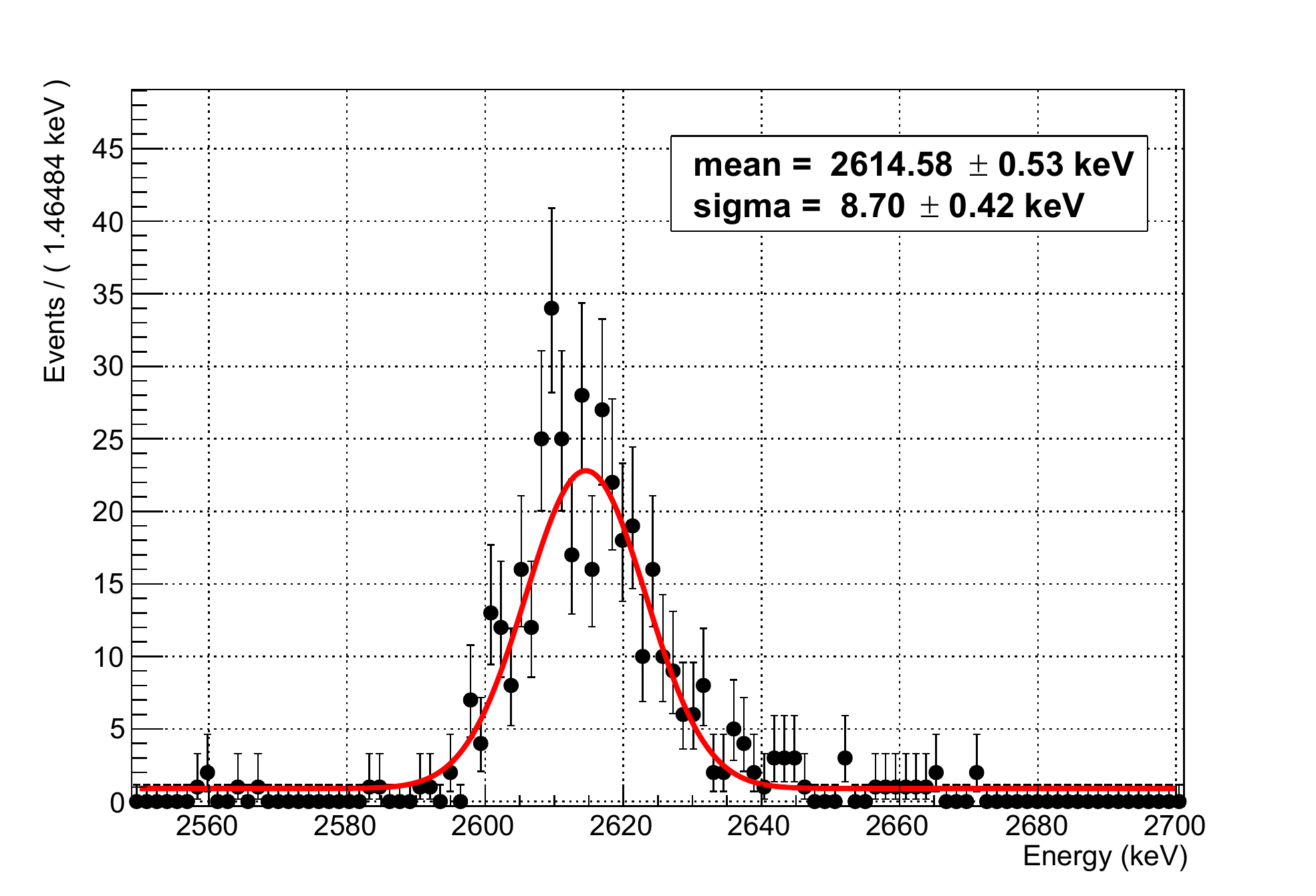}
\caption{Unbinned maximum likelihood fit of the 2615 keV line, modelled with a single gaussian with a constant background. 
}
\label{fig:Fit}       
\end{figure}

To find the rotation angle minimizing the resolution ($\theta_{\text{MIN}}$), different angles ($\theta$) are varied following an adequate minimization algorithm. For each $\theta$, the variable to be optimized is the full width at half maximum energy resolution (FWHM) evaluated at the 2615~keV peak with an unbinned maximum likelyhood fit, as shown in figure \ref{fig:Fit}. 

The resolution improvement will be then evaluated with the Resolution Gain (RG), defined by
\begin{equation}
\label{eq:Gain}
\text{RG}=\frac{\text{FWHM}(\theta_{\text{MIN}})}{\text{FWHM}(\theta=0)}\cdot 100.
\end{equation}
Once $\theta_{\text{MIN}}$ is obtained, the value of the RG parameter is calculated independently for each heat channel of CUPID-0 in order to check the goodness of the minimization procedure. RG is  evaluated for different energy peaks, to check whether if some energy dependence exists. Since the highest energy peak registered with the standard $^{232}$Th calibration is 2615~keV and the $^{82}$Se Q$_{\beta\beta}$ is $\sim$3000~keV, the analysis is conducted on the $^{56}$Co calibration, characterized by peaks above the $^{82}$Se Q$_{\beta\beta}$. By evaluating the Resolution Gain averaged on energy ($\overline{\text{RG}}$) on the $^{56}$Co calibration, this minimization method shows a clear independence from energy, alongside with the evaluation of the minimization effect both above and below Q$_{\beta\beta}$. 
\\
As previously stated, the resolution minimization has been performed using different methods, described in detail in the next sections.

\subsection{Single angle minimization} \label{sec:1D_MIN}

\begin{figure}[htbp]
\includegraphics[width=0.498\textwidth]{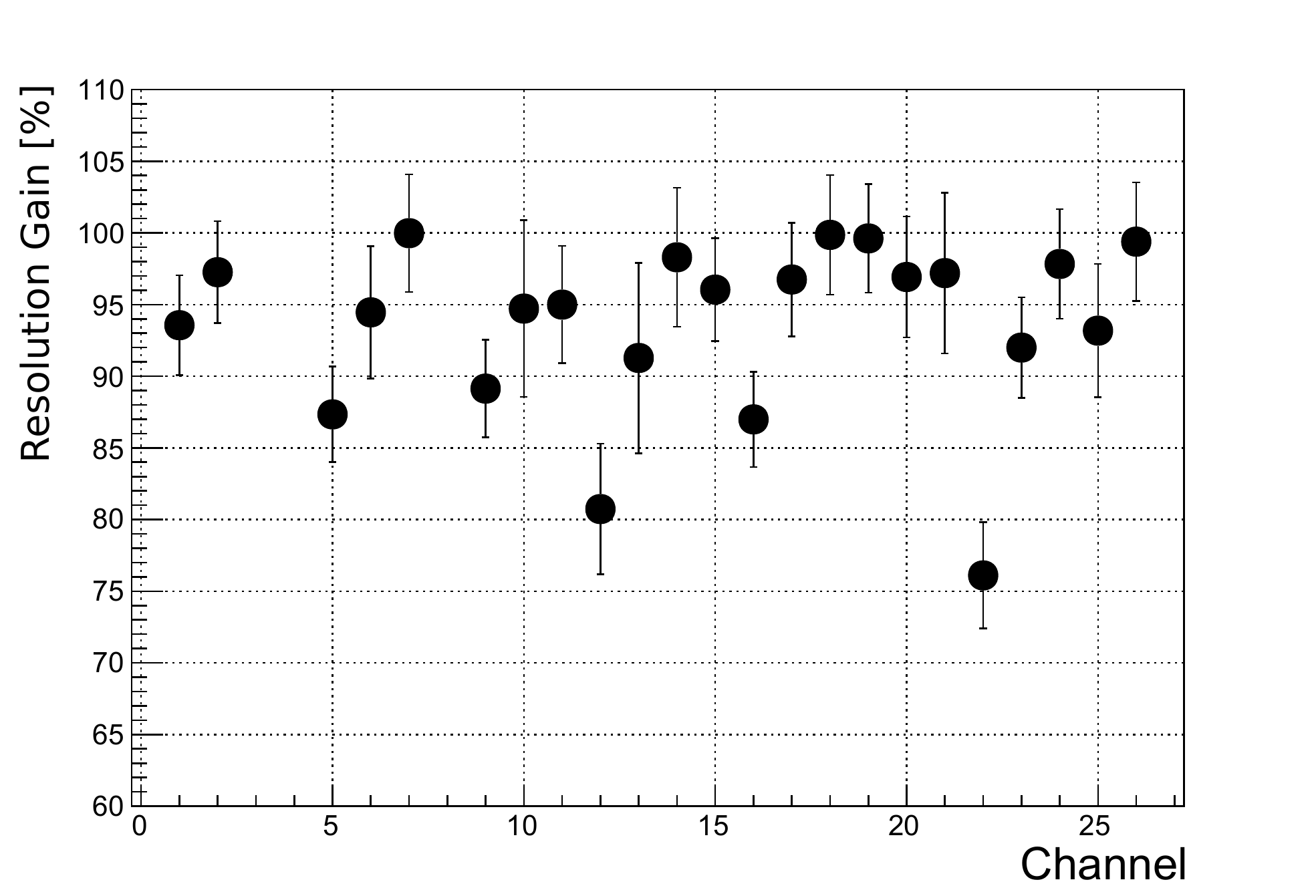}
\includegraphics[width=0.498\textwidth]{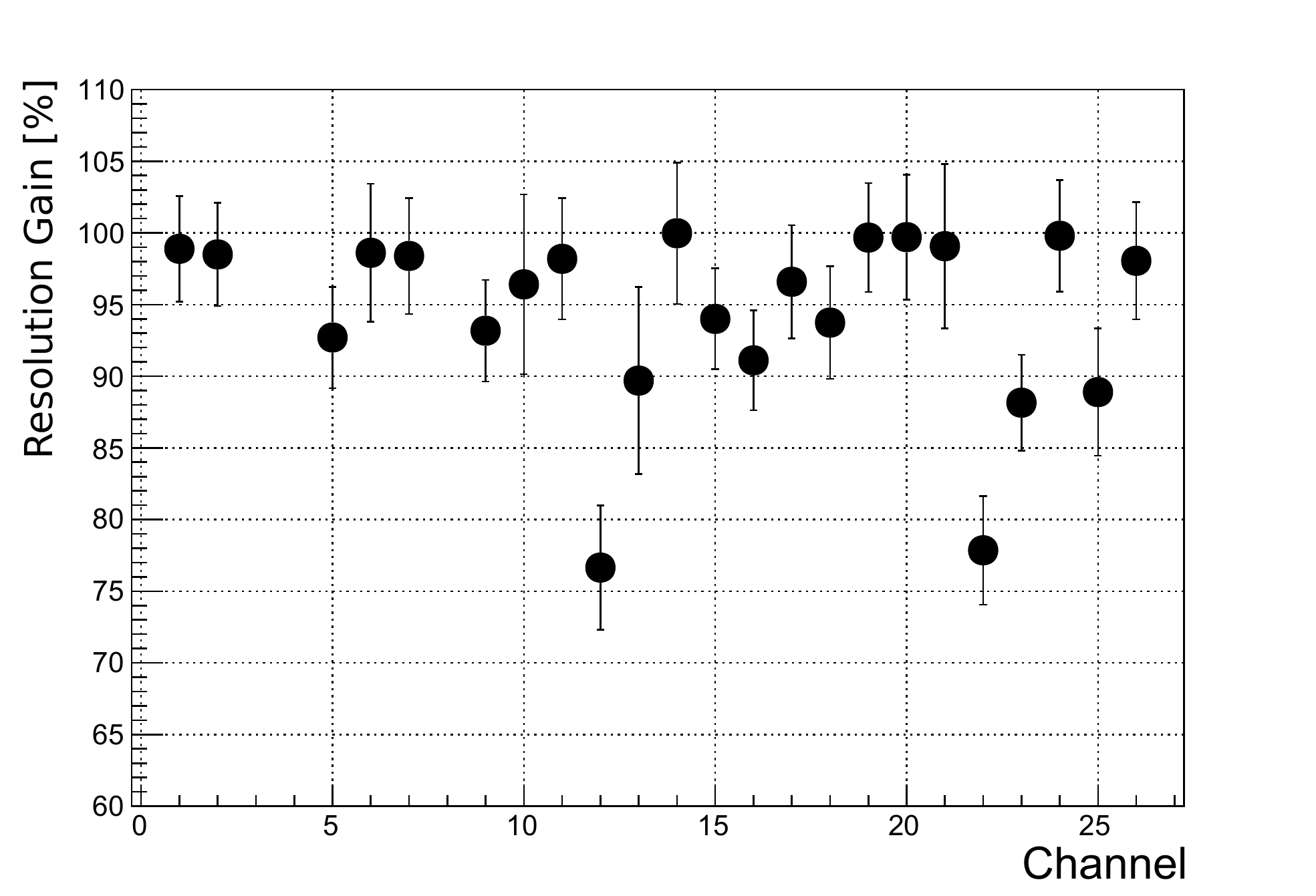}
\caption{Results for one dimensional minimization for all the heat channels in a single DS after Bottom (left) and Top (right) optimized rotations. The RG varies between channels with an average of $\sim$90~\%, depending on the strength of the correlation in each detector. The best results are obtained for channel 12 and channel 22.}
\label{fig:1D_CHRes}       
\end{figure}

To find $\theta_{\text{MIN}}$ for Top and Bottom light amplitudes, two separate one-dimension minimizations are performed. The chosen algorithm is the Golden-Section search, which successively narrows the range of values where the minimum is known to exist. This algorithm is efficient and guarantees a correct termination in $\sim$20 iterations. The results obtained performing independently the Top and Bottom best rotations for each channel are shown in figure \ref{fig:1D_CHRes}.

The obtained RG varies between channels depending on the strength of the correlation between light and heat amplitudes in each detector. The best results are obtained for channel 12 and channel 22. In particular channel 22 is the detector getting the higher benefit from this procedure, reaching $\text{RG}=(76\pm6)~\%$.

\begin{figure}[htb]
\includegraphics[width=0.498\textwidth]{DS1031_GAINBOT_NoLeg.pdf}
\includegraphics[width=0.498\textwidth]{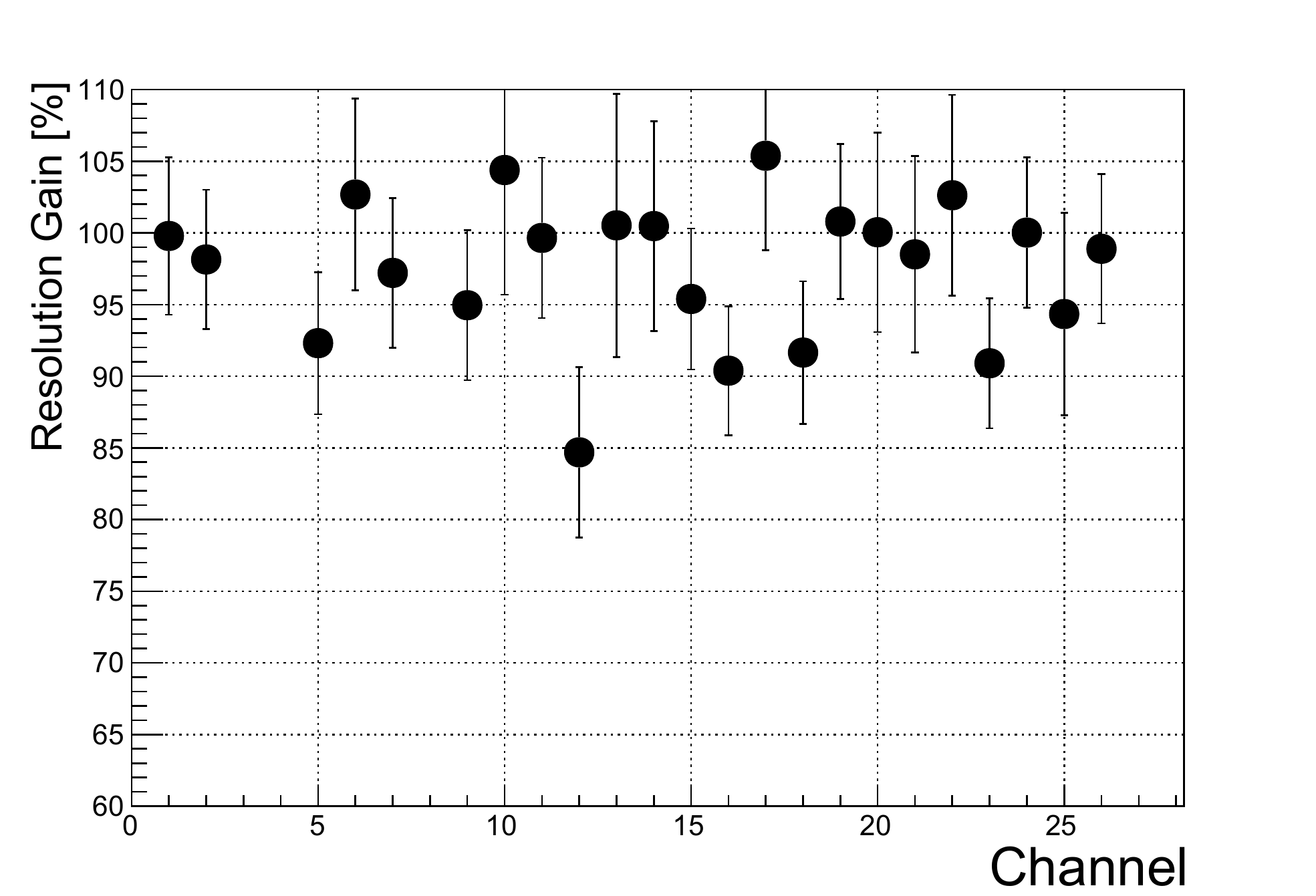}
\caption{
Results for one dimensional minimization for all the heat channels in a single DS with numerical (left) and analytical (right) minimization procedures. Both methods use a single light detector. The RG varies for the different channels, depending on the correlation's strength in each detector. The numerical method shows on average a better performance than the analytical one. The average improvement over all the channels is appreciably lower with this last method, and the uncertainties on each point are generally greater or equal. Considering for example channel 12, the two methods give both a better RG if compared to neighboring channels, but the numerical method gives a better result.}
\label{fig:AnVSNum_CHs}       
\end{figure}

The performance of this method are compared to the results obtained applying the analytical optimization of reference \cite{ZnSEDecorr} to the same set of data. The comparison is reported in figure \ref{fig:AnVSNum_CHs}. The numerical method shows on average a better performance than the analytical one. In particular RG for each channel are lower after the numerical minimization,
and the error bars on each point are generally smaller or equal. This behavior can be traced back to the aformentioned reasons, and provides a proof of the major reliability of the numerical method over the analytical one.

\begin{figure}[htbp]
\centering
\includegraphics[width=0.8\textwidth]{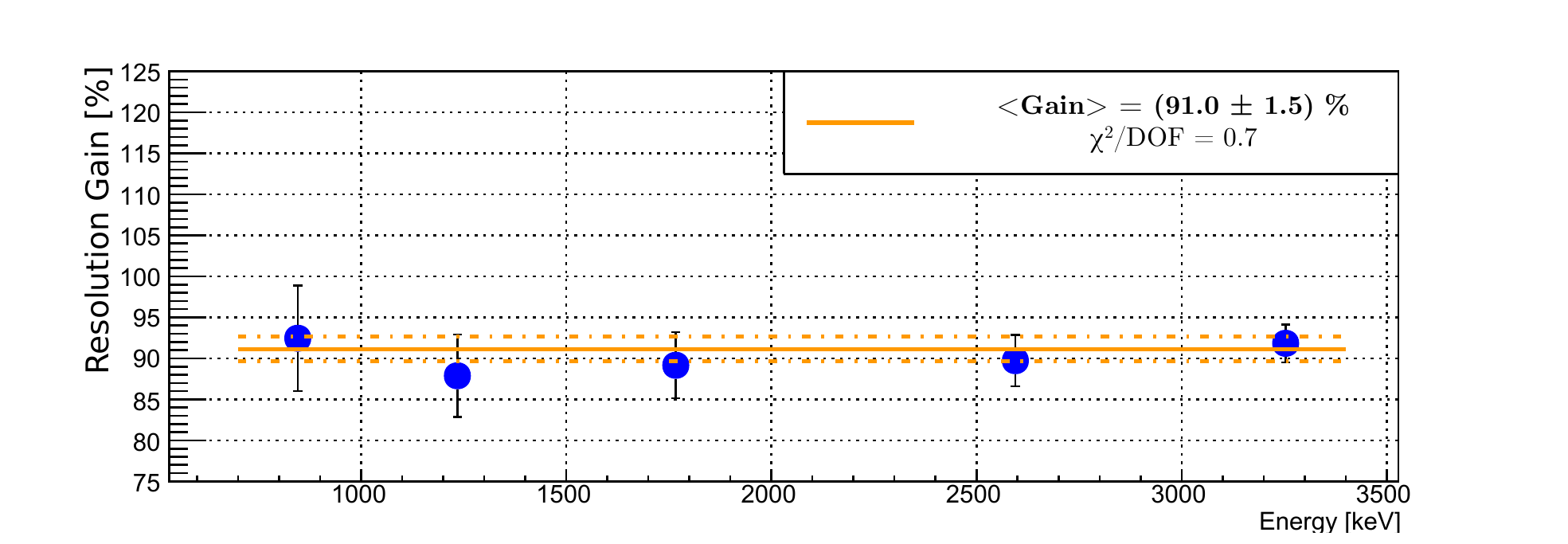}
\\
\includegraphics[width=0.8\textwidth]{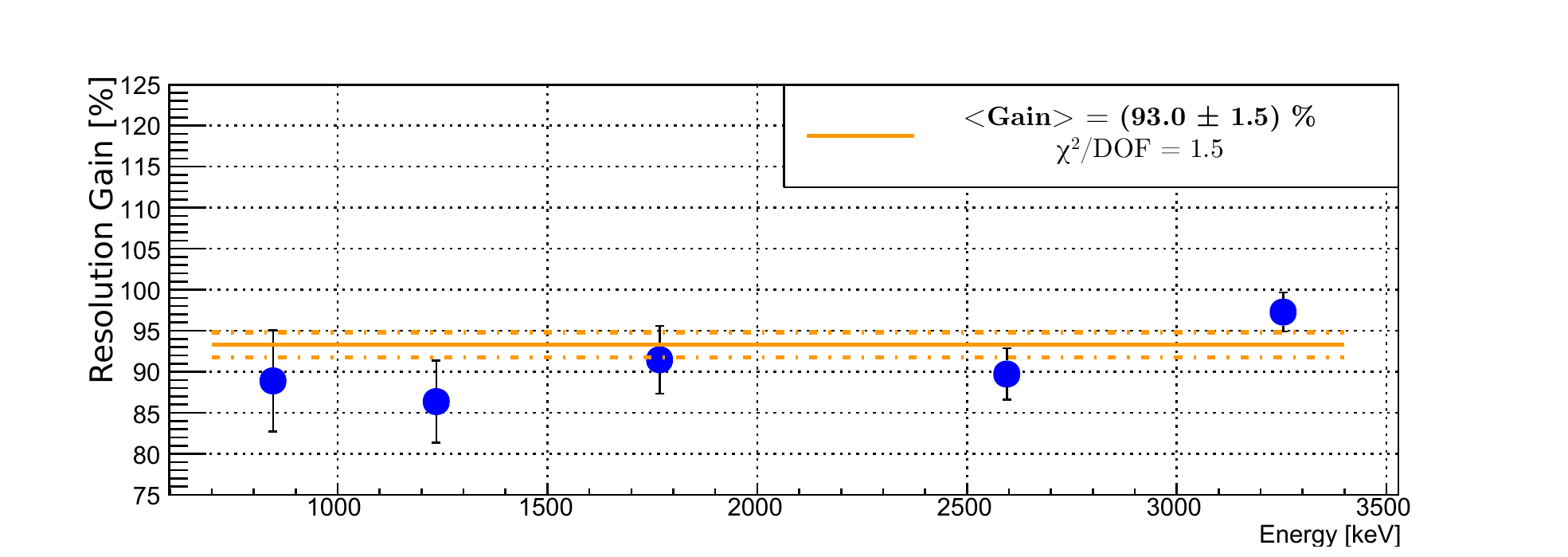}
\caption{$\overline{\text{RG}}$ evaluation for Bottom (up) and Top (low) decorrelations on $^{56}$Co calibration. In both cases the RG is constant with energy, proving the effectiveness of this technique both below and above $^{82}$Se $Q_{\beta\beta}$. The solid line represents the $\overline{\text{RG}}$ value, while the dashed lines delimit its the 68\% confidence interval.}
\label{fig:1D_CFRGainVSE_Co}       
\end{figure}

To investigate a possible dependency of RG on the energy, $\overline{\text{RG}}$ has been evaluated with the $^{56}$Co calibration over the (800-3000)~keV energy range. As shown in figure \ref{fig:1D_CFRGainVSE_Co}, the resolution enhancement affects all the energies, with the average values of $\overline{\text{RG}}_{\text{Top}}=(93.0\pm1.5)~\%$ and $\overline{\text{RG}}_{\text{Bottom}}=(91.0\pm1.5)~\%$. In both cases no energy dependence exists, proving that all possible $\beta$/$\gamma$ signal share the same correlation. 

\subsection{Double angle minimization}

Since the Top and Bottom light amplitudes are related to the same heat signal, a combined decorrelation has been performed taking into account these two informations simultaneously. The algorithm used to find the couple ($\theta_{\text{Bottom}},\theta_{\text{Top}}$) which minimizes the heat FWHM is the Nelder-Mead simplex method, iterated twice for each channel using two different starting points.
This strategy helps in dealing with channels characterized by different Top and Bottom correlations. For these detectors, both the Top and Bottom rotations have to be performed to minimize the FWHM, whose minimum cannot be reached exploiting only one light/heat correlation. A possible explanation resides in small differences in the emission spectrum of the various ZnSe crystals of the CUPID-0 detector. Such differences may cause a mismatch between crystal emission and ARC acceptance, resulting in a difference between measured Top and Bottom light amplitudes. The loss of light causes a different shape of the light/heat scatter plot for the two detectors, raising the need to perform different rotations for the two detectors.

\begin{figure}[htbp]
\centering
\includegraphics[width=0.5\textwidth]{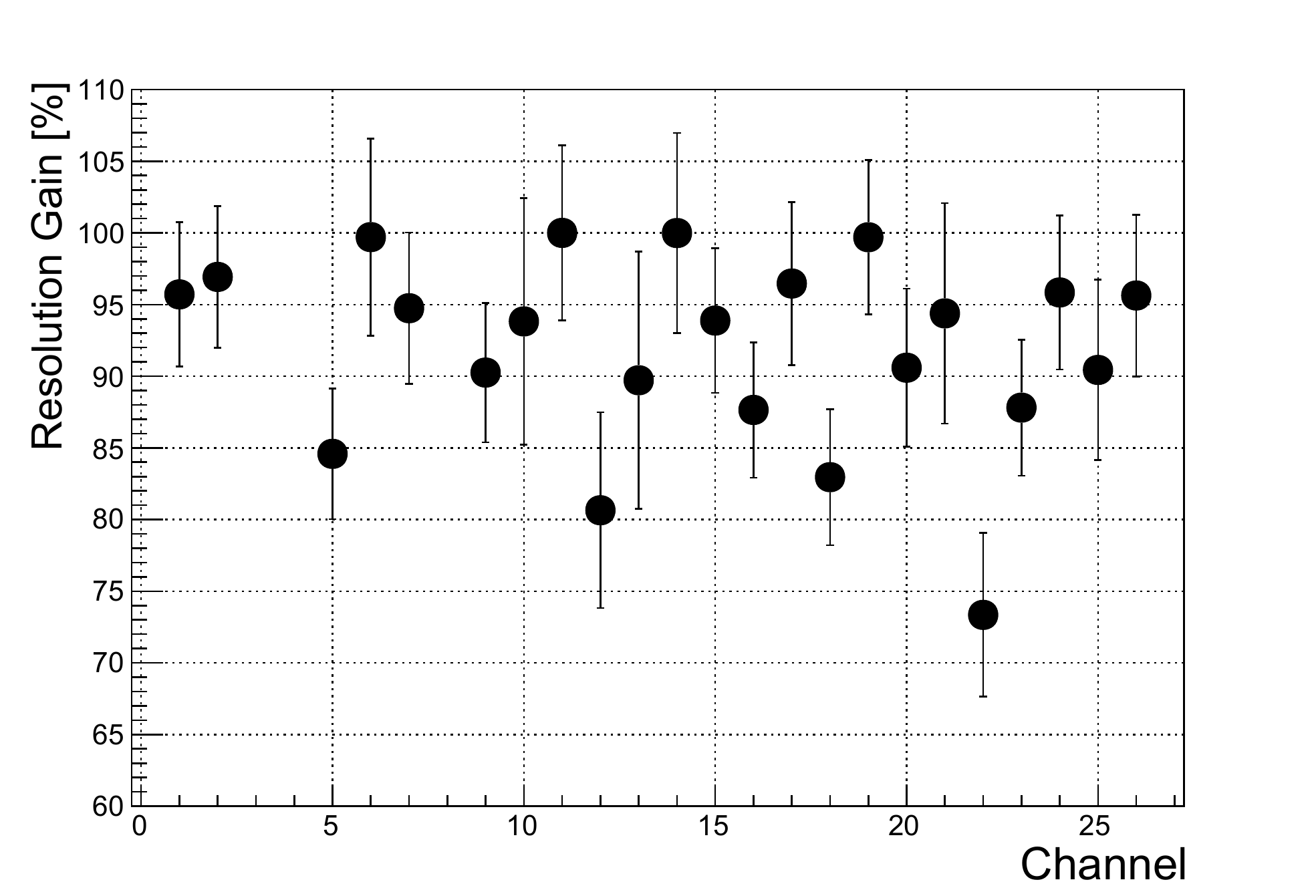}
\caption{$\overline{\text{RG}}$ obtained for each channel combining Top and Bottom light information. Comparing $\overline{\text{RG}}$ of channel 18 with the one reported in figure \ref{fig:1D_CHRes}, a better performance of the minimization algorithm can be appreciated.}
\label{fig:2D_CHRes}       
\end{figure}

The results of this new minimization strategy are reported, in terms of $\overline{\text{RG}}$s obtained for each channel, in figure \ref{fig:2D_CHRes}. The algorithm's capability in dealing with some channels can be appreciated comparing the new $\overline{\text{RG}}$ values with the ones reported in figure \ref{fig:1D_CHRes}. In particular, channel 18 has a marked improvement, being characterized by the aformentioned peculiar correlation. On the other hand, other channels, such as channel 11, are worsened by this second minimization algorithm. The cause of this worsening has to be searched in the performance of the chosen minimization algorithm, inefficient in dealing with flat minima regions. This result can be improved implementing better algorithms, but only few channels would benefit from such approach, making this effort unworthy of the increased complexity for our application.  

\begin{figure}[htbp]
\centering
\includegraphics[width=0.8\textwidth]{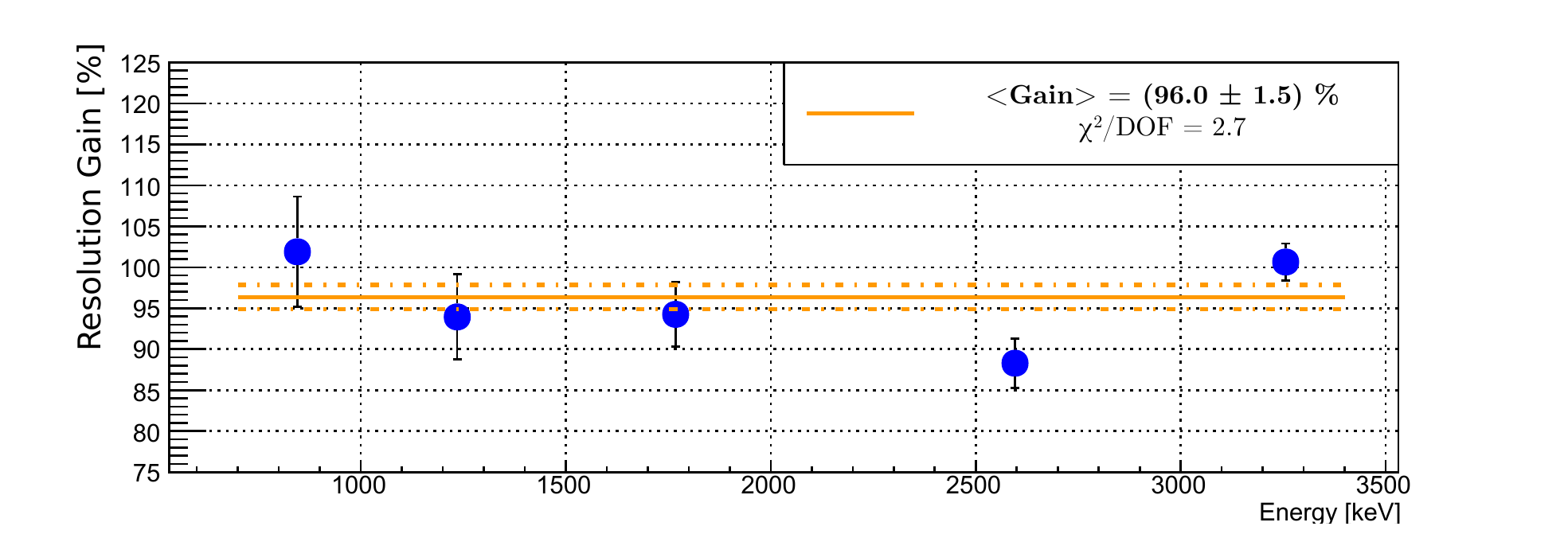}
\caption{$\overline{\text{RG}}$ evaluation for simultaneous Bottom and Top decorrelation on $^{56}$Co calibration. The obtained $\overline{\text{RG}}$ is worst with respect to the ones reported in figure \ref{fig:1D_CFRGainVSE_Co}, because of the difficulties the Nelder-Mead simplex algorithm has in finding a proper minimum. The solid line represents the $\overline{\text{RG}}$ value, while the dashed lines delimit its 68\% confidence interval.}
\label{fig:2D_CoRes}       
\end{figure}

The evaluation of $\overline{\text{RG}}$ over different energies with the $^{56}$Co calibration allows to evaluate the average performance of this algorithm. As reported in figure \ref{fig:2D_CoRes} an average $\overline{\text{RG}}=(96\pm1.5)~\%$ is obtained, significantly worsening the result obtained with the Bottom minimization, presented on the top of figure \ref{fig:1D_CFRGainVSE_Co}. The aformentioned difficulties of the Nelder-Mead simplex algorithm in finding a proper minimum are the main cause for this general worsening. As a direct consequence, the simultaneous combined minimization has to be considered only for those channels characterized by particular correlation characteristics.

\begin{figure}[htbp]
\centering
\includegraphics[width=0.5\textwidth]{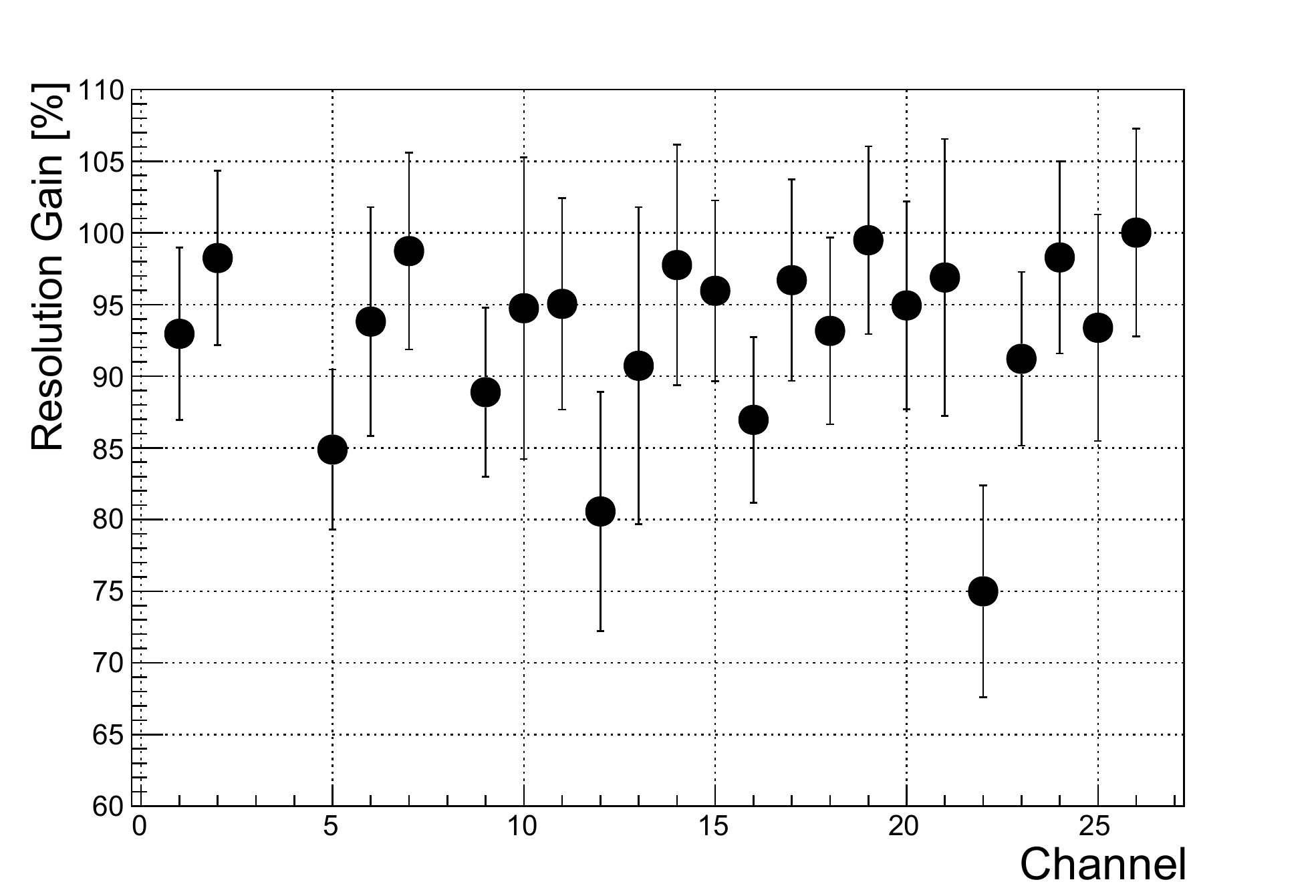}
\\
\includegraphics[width=0.8\textwidth]{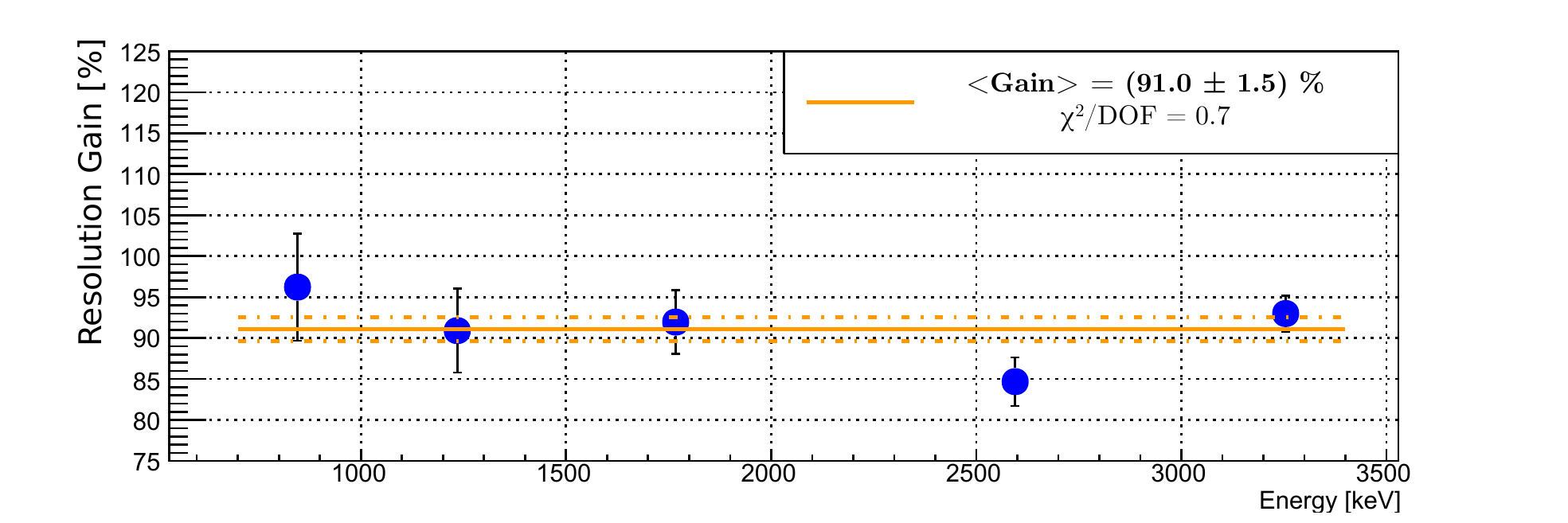}
\caption{$\overline{\text{RG}}$ evaluation with two subsequent rotations, the first in the Bottom light/heat plane and the second in the Top Light/Bottom corrected heat plane. The angle used are obtained with two distinct 1D minimization. In the upper panel the RG for each channel is shown. Comparing with figure \ref{fig:1D_CHRes} an improvement for channel 18 can be appreciated, as a consequence of the combined minimization. In the lower panel the $\overline{\text{RG}}$ evaluation on $^{56}$Co calibration is shown. A comparison with figure \ref{fig:1D_CFRGainVSE_Co} shows the compatibility between this method and the single angle rotation. The solid line represents the $\overline{\text{RG}}$ value, while the dashed lines delimit its 68\% confidence interval.}
\label{fig:2DMix_Result}       
\end{figure}

Given the above results, two conclusions can be extracted:
\begin{itemize}
    \item The minimization performed considering a single detector is effective in improving the resolution over the majority of the considered detectors. In particular the Bottom light/heat plane gives slightly better results on the average of all the channels, if compared to the Top light/heat plane.
    \item The minimization performed simultaneously on Top and Bottom light/heat plane has much better performances on single channels characterized by peculiar correlation features. 
\end{itemize}
To exploit both these features, a combined decorrelation has been performed by means of two subsequent one-dimensional rotations. Firstly $\theta_{\text{Bottom}}$ is found as described in section \ref{sec:1D_MIN}, then the Bottom rotation is fixed and a new rotation angle minimizing the resolution is found in the Top Light/Bottom corrected heat plane. The results, reported in figure \ref{fig:2DMix_Result}, show for the anomalously correlated channels (such as channel 18) the expected reduction and a $\overline{\text{RG}}$ value for the $^{56}$Co calibration compatible with the one obtained for the single Bottom minimization (see figure \ref{fig:1D_CFRGainVSE_Co}). The obtained result is aligned with the assumptions made for the definition of this analysis procedure, proving the consistency of the obtained results. 

\subsection{Chosen minimization method and total result}

\begin{figure}[htbp]
\centering
\includegraphics[width=0.49\textwidth]{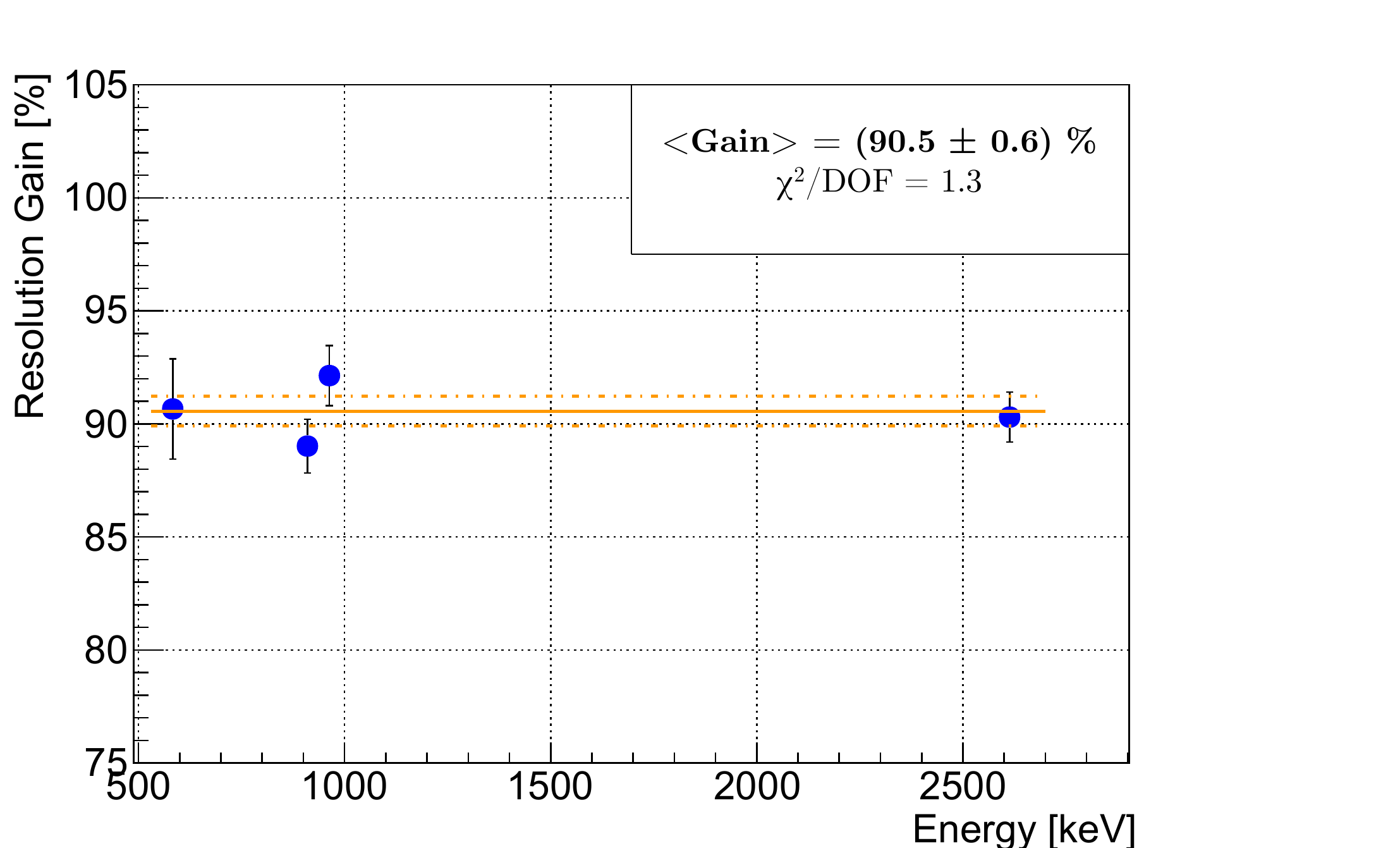}
\includegraphics[width=0.49\textwidth]{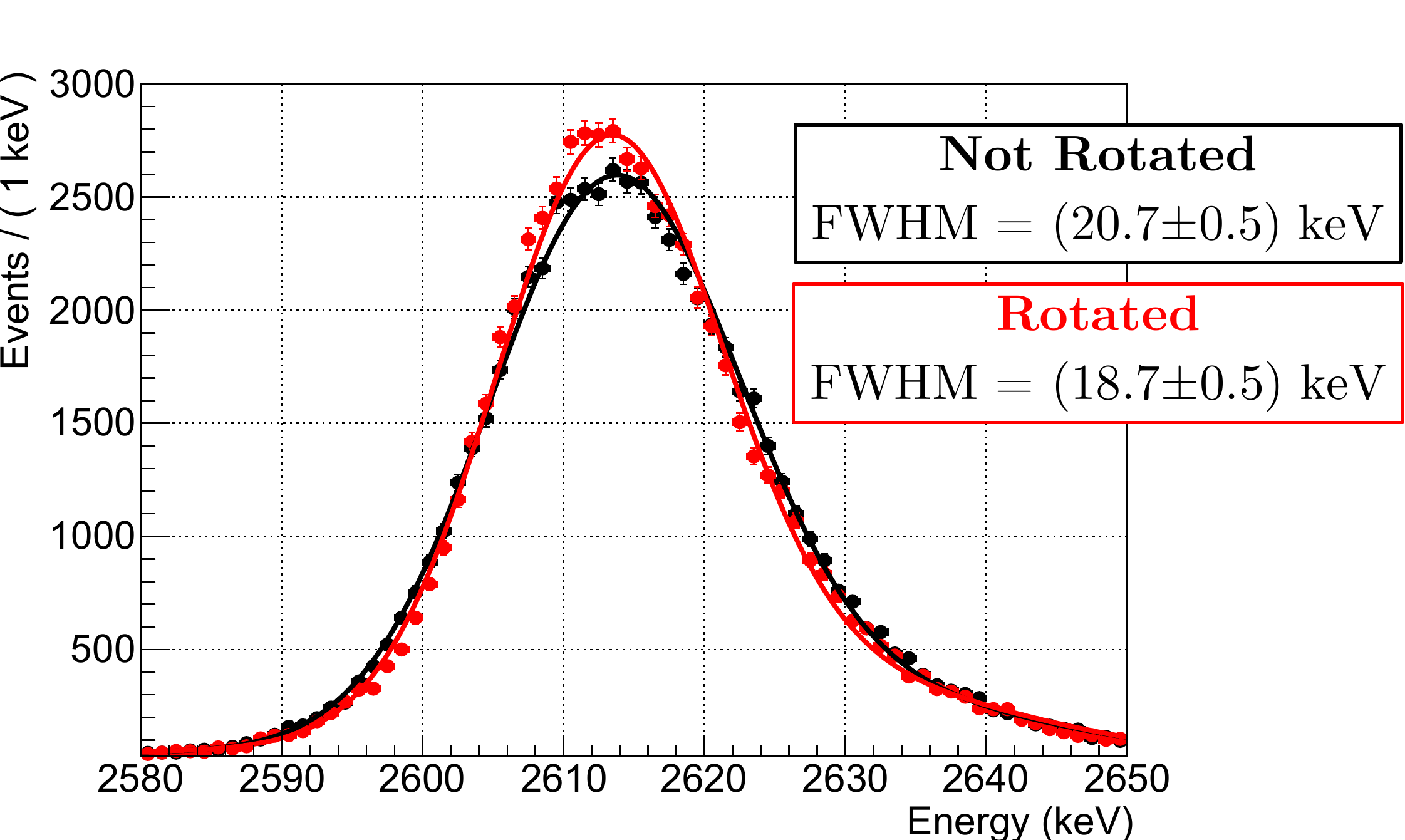}
\caption{Application of the best rotation method to the full calibration statistic of the CUPID-0 experiment. In the left panel the resolution gain over the energy is shown. The obtained gain is $\overline{\text{RG}}=(90.5\pm0.6)~\%$, corresponding to a change from $\text{FWHM}=(20.7\pm0.5)~\text{keV}$ to  $\text{FWHM}=(18.7\pm0.5)~\text{keV}$.  The solid line represents the $\overline{\text{RG}}$ value, while the dashed lines delimit its 68\% confidence interval. In the right panel the shape of the 2615~keV peak is shown before and after the rotation. The main effect of the rotation is the narrowing of the peak, while the line shape is not affected by the rotation procedure.}
\label{fig:GlobalResults_CAL}  
\end{figure}

Comparing all obtained results, the best method to select the optimal rotation angle is to apply two subsequent one-dimensional rotations. This procedure allows to preserve the improvement obtained with the single rotation, while exploiting the particular features of some channels. In figure \ref{fig:GlobalResults_CAL} the effect of the combined rotation is shown on the whole $^{232}$Th calibration statistics acquired by CUPID-0. An average $\overline{\text{RG}}=(90.5\pm0.6)\%$ is obtained over all the energies, corresponding to a change from $\overline{\text{FWHM}}=(20\pm1)~\text{keV}$ to  $\overline{\text{FWHM}}=(18\pm1)~\text{keV}$ at the 2615~keV peak.  

\section{Conclusion}
Correcting the light/heat positive correlation of ZnSe scintillating bolometers has proven to be an efficient method to improve the resolution in the CUPID-0 experiment. The correction has been implemented with a rotation in the light/heat amplitude plane, selecting the angle giving the best resolution gain. Three different methods have been investigated to select the best angle: a resolution minimization considering Top and Bottom light independently, a combined minimization using the two amplitudes simultaneously, and two subsequent one-dimensional minimizations. The first method showed good energy stability over a wide energy range, proving the performance and the applicability of this strategy. On the other hand, in the case of crystals characterized by a peculiar light/heat correlation, the minimization algorithm considering simultaneously the two LDs has to be preferred. The obtained resolution gain highly increases for the single channel, while keeping the positive features of the single angle minimization. As a consequence, the optimization using two subsequent one-dimensional minimization has been selected as the general processing method, since better results could be obtained with this method with respect to using the Top amd Bottom light amplitude simultaneously. An overall $\overline{\text{RG}}=(90.5\pm0.6)\%$ is obtained on the whole CUPID-0 calibration statistics, meaning a reduction from $\text{FWHM}=(20.7\pm0.5)~\text{keV}$ to  $\text{FWHM}=(18.7\pm0.5)~\text{keV}$ at the 2615~keV peak. Such an improvement can contribute in pushing the sensitivity for $0\nu\beta\beta$ search in CUPID-0 and other rare events physics analysis \cite{EXCITED_STATES}.

\bibliographystyle{JHEP}      
\bibliography{references.bib}

\end{document}